\documentclass[12pt,twoside]{article}
\usepackage{fleqn,espcrc1}
\usepackage{multicol}
\usepackage[dvips]{epsfig}

\title{Meson Production in p+d Reactions}
\author{GEM Collaboration\\ M. Betigeri$^{\ i}$, J. Bojowald$^{\ a}$, A. Budzanowski$^{\ d}$, A.
Chatterjee$^{\ i}$, J. Ernst$^{\ g}$, L. Freindl$^{\ d}$, D.
Frekers$^{\ h}$, W. Garske$^{\ h}$, K. Grewer$^{\ h}$, A. Hamacher$^{\
a}$, J. Ilieva$^{\ a,e}$, L. Jarczyk$^{\ c}$, K. Kilian$^{\ a}$, S.
Kliczewski$^{\ d}$, W. Klimala$^{\ a,c}$, D. Kolev$^{\ f}$, T.
Kutsarova$^{\ e}$, J. Lieb$^{\ j}$, H. Machner$^{\ a}$, A. Magiera$^{\
c}$, H. Nann$^{\ a}$, L. Pentchev$^{\ e}$, H. S. Plendl$^{\ k}$, D.
Proti\'c$^{\ a}$, B. Razen$^{\ a}$, P. von Rossen$^{\ a}$, B. J.
Roy$^{\ i}$, R. Siudak$^{\ d}$, A. Strza{\l}kowski$^{\ c}$, R.
Tsenov$^{\ f}$, K. Zwoll$^{\ b}$\\ \small \noindent {\it a. Institut
f\"{u}r Kernphysik, Forschungszentrum J\"{u}lich, J\"{u}lich, Germany}\\ \small
\noindent {\it b. Zentrallabor f\"ur Elektronik, Forschungszentrum
J\"ulich, J\"ulich, Germany}\\ \small \noindent {\it c. Institute of
Physics, Jagellonian University, Krakow, Poland}\\ \small \noindent
{\it d. Institute of Nuclear Physics, Krakow, Poland}\\ \small
\noindent {\it e. Institute of Nuclear Physics and Nuclear Energy,
Sofia, Bulgaria}\\ \small \noindent {\it f. Physics Faculty, University
of Sofia, Sofia, Bulgaria}\\ \small \noindent {\it g. Institut f\"ur
Strahlen- und Kernphysik der Universit\"at Bonn, Bonn, Germany}\\
\small \noindent {\it h. Institut f\"ur Kernphysik,  Universit\"at
M\"unster, M\"unster, Germany}\\ \small \noindent {\it i. Nuclear
Physics Division, BARC, Bombay, India}\\ \small \noindent {\it j.
Physics Department, George Mason University, Fairfax, Virginia, USA}\\
\small \noindent {\it k. Physics Department, Florida State University,
Tallahassee, Florida, USA} }
\begin{document}
\maketitle

\section{Introduction}
The deuteron is a loosely bound system with large distance between the
two nucleons. It seems therefore a good testing ground for the impulse
approximation. In addition, its wave function as well as those of the
produced light nuclei are believed to be well known and one can hope
that a theoretical treatment in the three nucleon sector might be
possible. The theoretical attempts in $pd\to (A=3)\ meson$ have not
been particular successful although a lot of effort was devoted to this
subject over the years \cite{Can98}. Similarly, the data base is by far
not complete and sometimes contradictory. We have, therefore, measured
the reactions $pd\to{^3He}\eta$ in the region of the N*(1535) resonance
and the reactions $pd\to{^3He}\pi^0$ and $pd\to{^3H}\pi^+$ below and in
the range of the $\Delta (1232)$ resonance.

\section{Experiments}
\begin{figure}[h]
\epsfig{figure=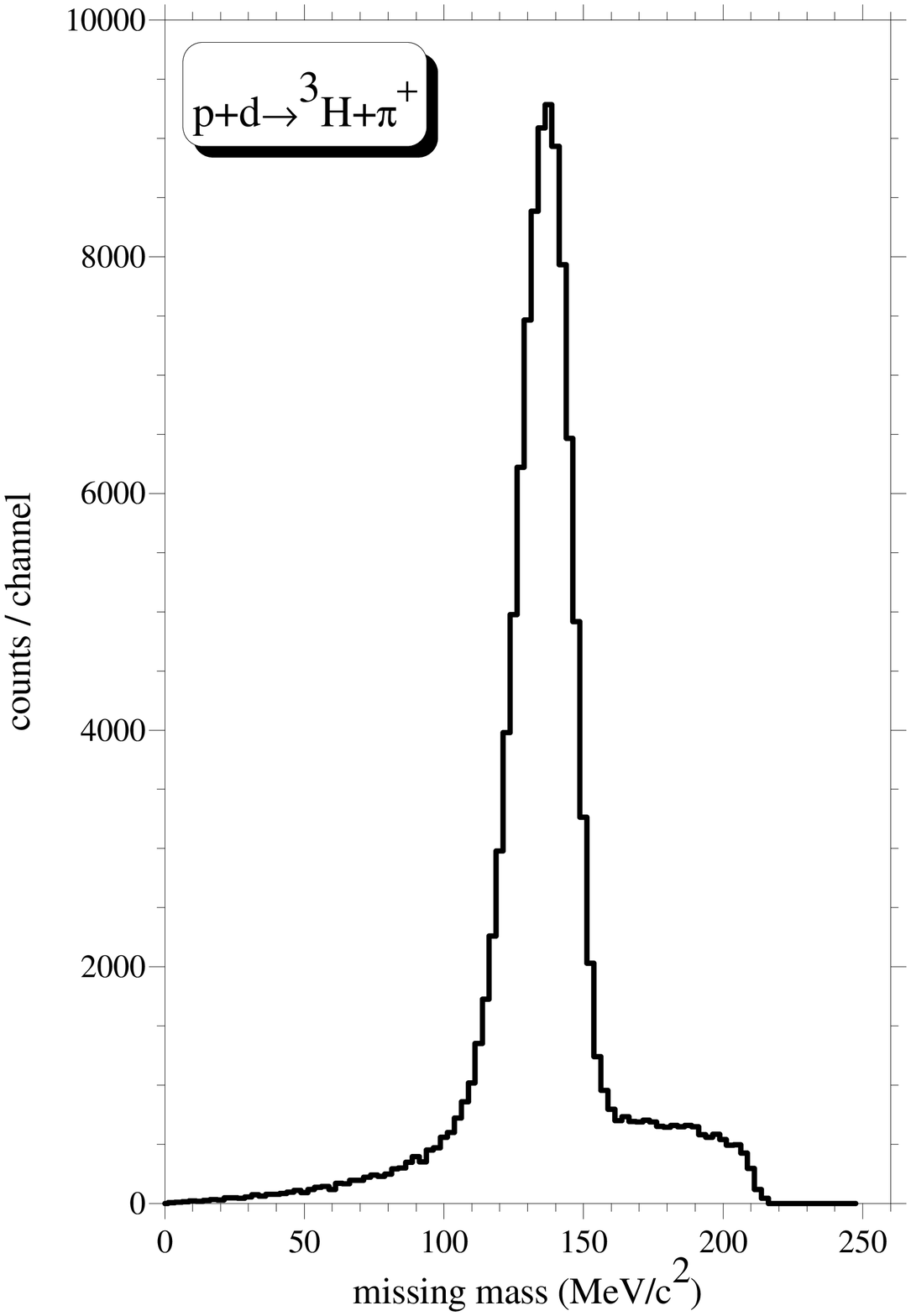, width=7cm, angle=0}
\epsfig{file=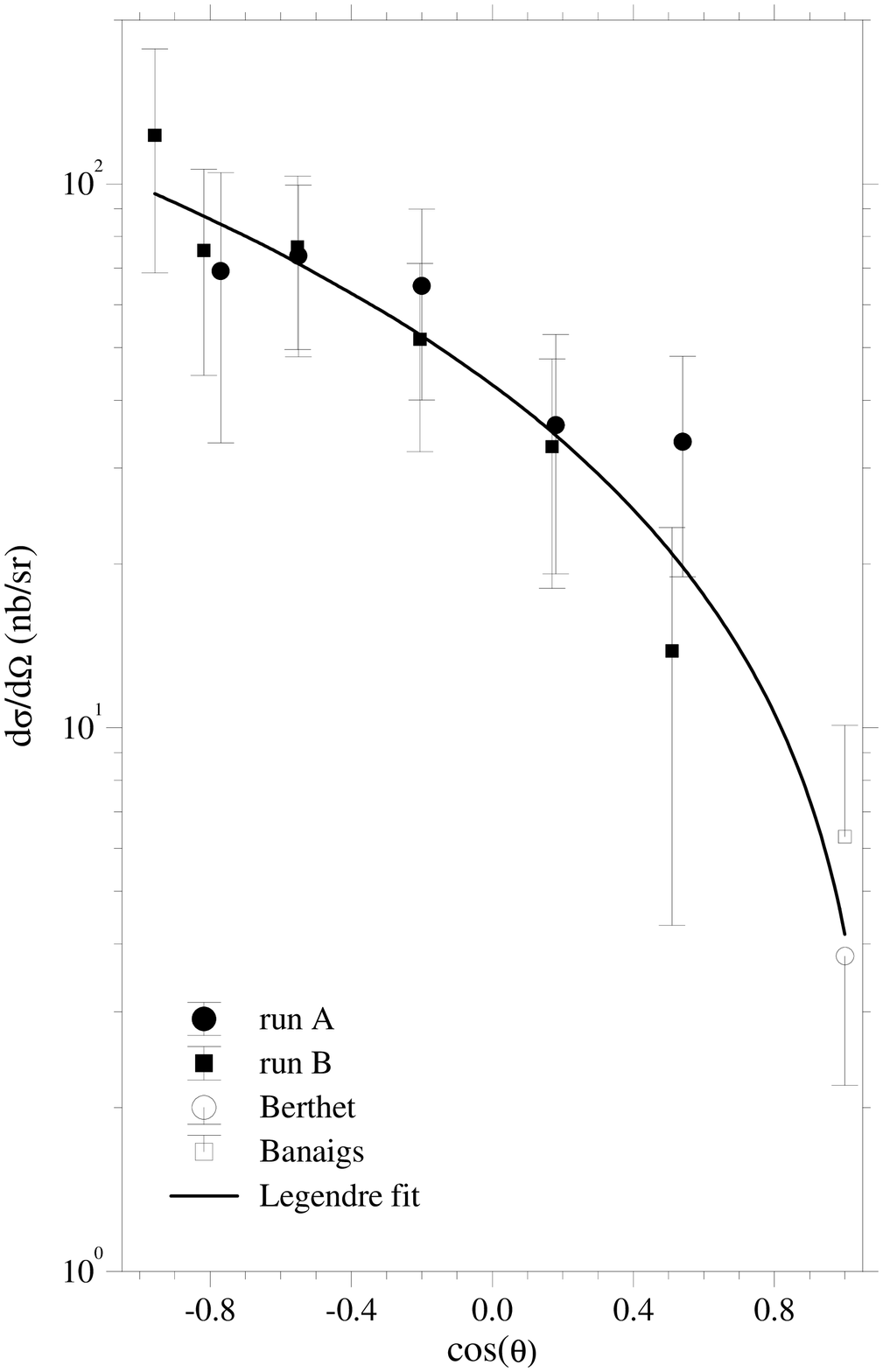,width=7cm} 
\begin{multicols}{2}
\caption{\label{mm}Deduced missing mass for the charged pion from the
kinematical complete measurement of the triton. The small background
was subtracted for each angular bin. } \caption{\label{eta} Angular
distribution of $\eta$ production in the cm system. Two points from the
indicated references are added (Ref. \cite{Ban73,Ber85}). The solid
curve indicates a Legendre polynomial fit.}
\end{multicols}
\end{figure}

The proton beams were extracted from the cooler-- synchrotron COSY in
J\"{u}lich. Although the beam was not cooled it had typically emittance of
$2.5 \pi\ mm\ mrad$ in all directions. The detector used is a stack of
diodes made from high purity Germanium. The diodes have structures on
the front side as well as on the rear side allowing track
reconstruction, energy measurement and particle identification. With
this detector call Germanium Wall Ref. \cite{Bet99} we measured the
heavy $A=3$ recoils. Together with conservation laws the four momentum
vectors of the unobserved mesons could be extracted. Recoils being
emitted under zero degree in the laboratory were measured with the
magnetic spectrograph BIG KARL. Both detector elements together form
the GEM detector. In Fig. \ref{mm} the missing mass spectrum for
charged pions from the $pd\to{^3H}\pi^+$ reaction at 850 MeV/c is
shown. Two things are worth mentioning: the high statistics in the
experiment and the small background.

\section{Results}
\subsection{$\eta$ Production} For the $\eta$-production we obtained an angular
distribution \cite{Bet00} which is dominated by p-wave (see Fig.
\ref{eta}) in contrast to near threshold, where the angular
distributions are s-wave dominated. A Legendre polynomial of second
order was found to account for the data. From this fit a total cross
section of $\sigma =573 \pm 83\ (stat.) \pm 69\ (syst.)\ nb$ was
deduced. This result is close to the one obtained by Banaigs et al.
\cite{Ban73} at a slightly higher energy. All cross section from
threshold up to $T_p\approx 1500\ MeV$ could be nicely accounted by a
simple calculation. This energy region corresponds to the centre of the
$N^\ast$ $S_{11}$ resonance ($\Gamma$ $\sim$ 200 MeV) known to couple
strongly to the $\eta$--$N$ channel \cite{PDG99}. One may therefore
attempt to describe the cross section by an intermediate $N^\ast$(1535)
resonance excitation:
\begin{equation}\label{cross_sect}
  \sigma(E)=\frac{p_\eta}{p_p}|M(E)|^2
\end{equation}
with $E$ the excitation energy and $M$ the matrix element which is
calculated as in photoproduction on the proton \cite{krusche} as
Breit--Wigner form with an energy dependent width. All parameters were
taken from Ref.'s \cite{krusche} and \cite{PDG99}. The only free
parameter is the strength fitted to the present data point. The trend
of the data is reproduced, which may be taken as an indication that
production of the $N^\ast$(1535) resonance is the dominant reaction
mechanism and that the product of kinematics and form factor changes
only very little over the present energy range. Deviations occur near
threshold which might be an indication of strong final state
interactions.

\subsection{$\pi$ Production}
\begin{figure}[h]
\epsfig{file=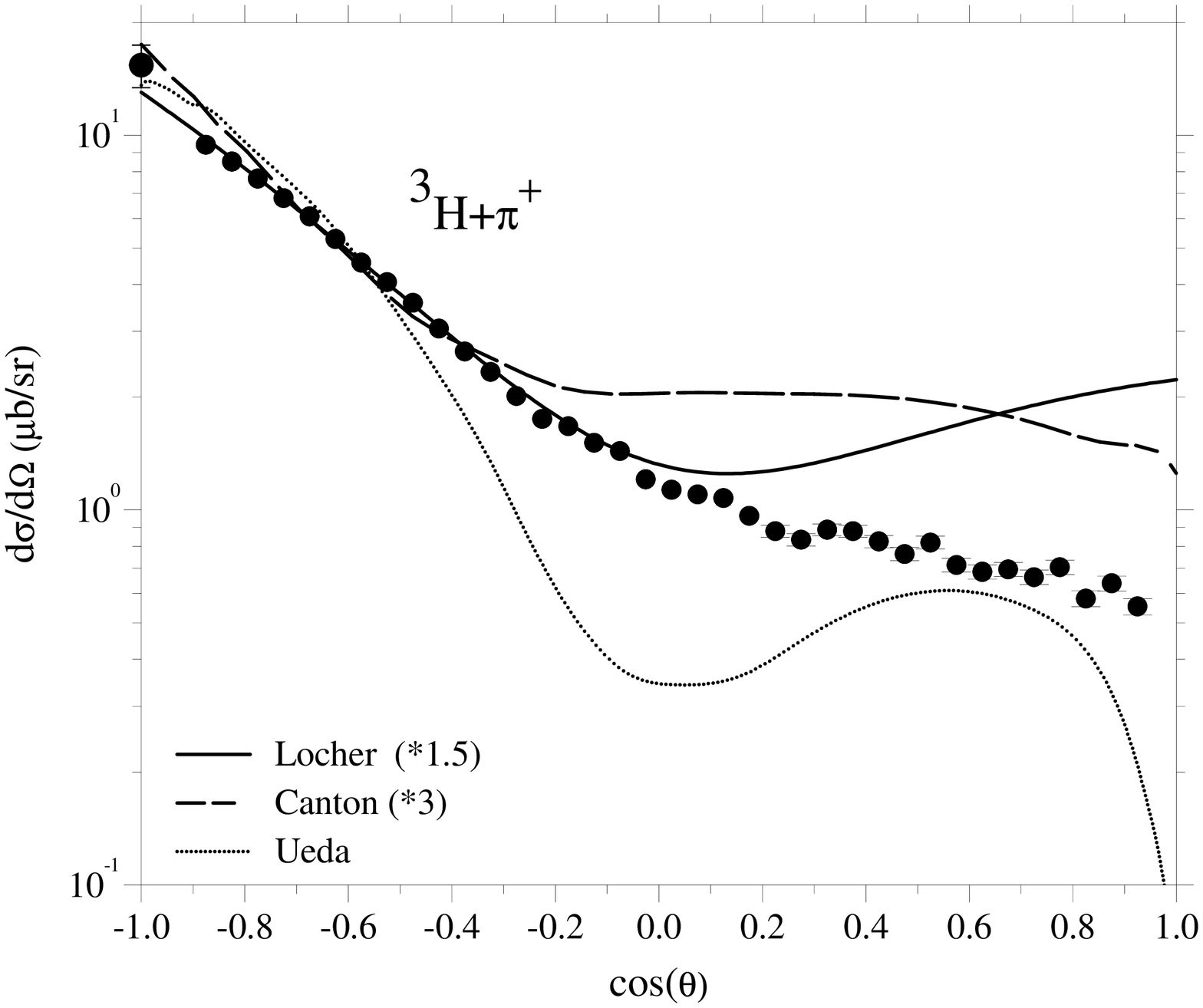,height=6cm}
\epsfig{file=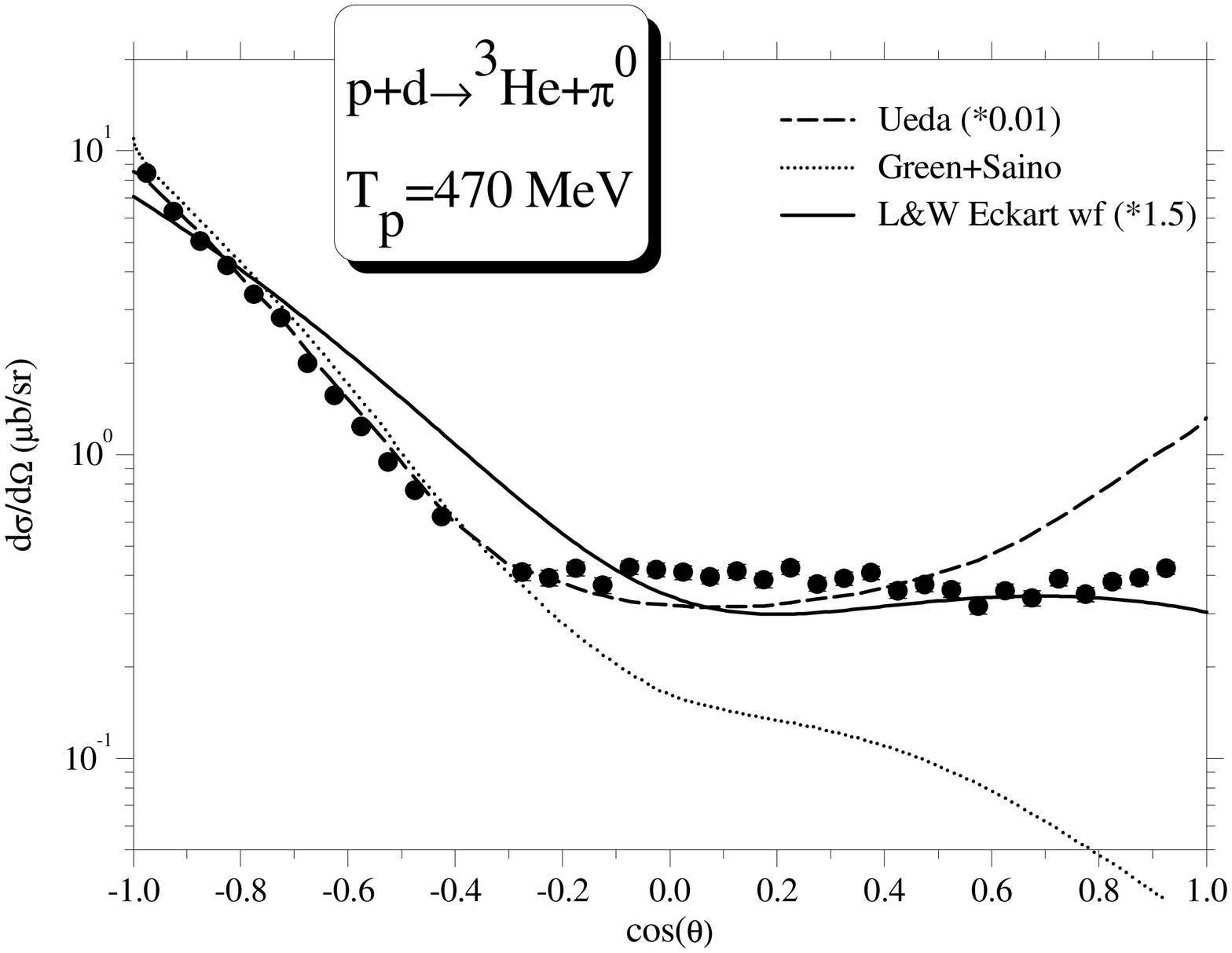,height=6cm} 
\begin{multicols}{2}
\caption{\label{pion_850}  Angular distribution for pion  emission for
the reaction $pd\to{^3H}\pi^+$ at a beam momentum of 850 MeV/c. The
curves are calculations from Ref.'s \cite{LW74,Ueda,Can98}.}
\caption{\label{pion_1050}Same, but for $pd\to{^3He}\pi^0$ emission at
a beam momentum of 1050 MeV/c. The calculations are according to Ref.'s
\cite{LW74,Ueda,Saino}. }
\end{multicols}
\end{figure}

For the $\pi$-production the situation for the angular distributions
very close to threshold is similar: they are isotropic. For higher beam
momenta the angular distribution becomes again backward peaked for the
heavy recoil. For a beam of 750 MeV/c it has an almost exponential
slope. For higher momenta an isotropic component shows up with
increasing importance with increasing beam momentum. As examples the
angular distributions for charged pion production at 850 MeV/c  and
neutral pion production at 1050 MeV/c are shown in Fig.'s
\ref{pion_850} and \ref{pion_1050}, respectively. One point measured at
$\cos(\theta)=-1$ which corresponds to zero degree in the laboratory
was measured with the magnetic spectrograph BIG KARL.

As examples the data are compared with several model calculations also
shown in Fig.'s  \ref{pion_850} and \ref{pion_1050}. Here, we restrict
ourselves to comparisons with published calculations for beam momenta
close to the present ones or perform such calculations in the very
transparent Locher--Weber model \cite{LW74}. The differential cross
section in this model is given by
\begin{equation} \label{Locher}
\frac{{d\sigma }}{{d\Omega }}\, = \,S\,\,K\,|F_D (q) - F_E (q)|^2
\,\frac{{d\sigma }}{{d\Omega }}(pp \to d\pi ^ +  )
\end{equation}
with $S$, a spin factor, $K$ a kinematical factor, and $F_D$, the
direct form factor and $F_E$, the exchange form factor, i.e. an elastic
$\pi d$ scattering after pion emission from the incident proton. The
graph corresponding to $F_D (q)$ is treated by most calculations. The
form factors were evaluated with emphasis on the short
range--components of the deuteron and the triton. This is achieved by
fitting the free parameters in a Hulth\'{e}n function to the
deuteron--charge form factor and similarly for tritium, the parameters
for the Eckart function and the 3-pole function to the tritium--charge
form factor. Also other functional dependencies taken from the work of
Fearing \cite{Fea77} were tried applying the same normalization as in
the work of Locher and Weber \cite{LW74}. All calculations need
normalization factors when compared to the data. Best results were
obtained for the Eckart wave function and an exponential (see Fig.'s
\ref{pion_850} and \ref{pion_1050}). We also compare the data with
calculations from Ueda \cite{Ueda}. The normalization factor is 0.01.
For details we refer to the original work. The local minimum at $\cos
(\theta)\approx 0$ for 850 MeV/c as well as the strong rise at forward
angles for 1050 MeV/c is not supported by the data. This range is
rarely reproduced by a calculation. Green and Saino \cite{Saino} even
claim their calculations to be unreliable in this range. The large and
almost exponential shaped component may be attributed to the direct
production graph. Therefore, new physics may be hidden in the small
more isotropic component.


\newcommand{\etal}{{\em et al.}}

%


\begin{thebibliography}{99}
\bibitem{Can98} L. Canton, G. Cattapan, G. Pisent, W. Schadow, J. P.
Svenne, Phys. Rev. {\bf C 57} (1998) 1588; L. Canton and W. Schadow,
Phys. Rev: {\bf C 56} (1997) 1231
\bibitem{Bet99} M. Bettigeri \etal, Nucl. Instrum. Methods Phys. Res. {\bf A421}, 447
(1999).
\bibitem{Bet00} M. Bettigeri \etal, Phys. Lett. {\bf B472}, 267
(2000).
\bibitem{Ban73} J. Banaigs \etal, Phys. lett. {\bf B45}, 394 (1973).
\bibitem{Ber85} P. Berthet \etal, Nucl. Phys. {\bf C443}, 589 (1985).
\bibitem{PDG99} Particle Data Group: C. Caso etal., The European Phys. J. 3 (1999) 1.
\bibitem{krusche} B. Krusche, Acta Phys. Polonica 27B (1996) 3147.
\bibitem{LW74} M. P. Locher and H. J. Weber, Nucl. Phys. {\bf B76}, 400 (1974).
\bibitem{Fea77} H. W. Fearing, Phys. Rev. {\bf C 16} (1977) 313.
\bibitem{Ueda} T. Ueda, Nucl. Phys. {\bf A505}, 610 (1974).
\bibitem{Saino} A. M. Green and M. E. Sainio, Nucl. Phys. {\bf A 329}
(1979) 477.
\end{thebibliography}
\end{document}